\documentclass[11pt]{article}
\UseRawInputEncoding
\usepackage{acl}
\usepackage{tikz}
\usepackage[most]{tcolorbox}
\usepackage{enumitem}
\usepackage{times}
\usepackage{latexsym}

\usepackage[T1]{fontenc}
\usepackage{url}
\usepackage[utf8]{inputenc}
\usepackage{graphicx}
\usepackage{color}
\usepackage{tabularx}
\usepackage{tabu}
\usepackage{booktabs}
\usepackage{multirow}
\usepackage{placeins}
\usepackage{graphicx} 
\usepackage{float}
\usepackage{subfigure} 
\usepackage{amssymb}
\usepackage[table]{xcolor} 
\usepackage{hyperref}
\usepackage{tablefootnote}
\usepackage{xspace}
\usepackage{arydshln}
\usepackage{xcolor}
\newcommand{\Corr}{{\small\fcolorbox{red}{white}{\textcolor{red}{Corr}}}}
\newcommand{\Cite}{{\small\fcolorbox{blue}{white}{\textcolor{blue}{Cite}}}}
\usepackage{fontawesome} 
\usepackage{tcolorbox}
\tcbuselibrary{listings,breakable}
\usepackage{listings}
\newtcblisting{jsonbox}{
  listing only,
  breakable,
  colback=gray!5!white,
  colframe=gray!75!black,
  listing options={basicstyle=\ttfamily\footnotesize,breaklines=true}
}
\usepackage{algorithm,algorithmicx}
\floatname{algorithm}{} 
\usepackage{float}
\usepackage{amsmath}
\usepackage{bm}
\usepackage{algorithmicx}
\usepackage{algorithm}
\usepackage{algpseudocode}
\usepackage{tikz}
\usetikzlibrary{shapes.geometric, arrows.meta, positioning}
\usepackage{enumitem}
\usepackage{arydshln}
\usepackage{array}
\renewcommand{\arraystretch}{0.9}
\usepackage{amssymb}
\usepackage{pifont}
\usepackage{enumitem}
\usepackage[most]{tcolorbox}
\usepackage{lipsum}
\usepackage{microtype}
\usepackage{caption}  

\newcommand{\ours}{\textsc{SciRAG}\xspace}
\newcommand{\eg}{\hbox{\emph{e.g.,}}\xspace}

\usepackage{svg}

\usepackage{titlesec}

\definecolor{SJTUred}{RGB}{180,0,32} 
\definecolor{YaleBlue}{RGB}{0,53,204} 
\definecolor{NYUPurple}{RGB}{140,40,200}
\definecolor{TataBlack}{RGB}{0,0,0} 
\definecolor{AIpink}{RGB}{255,105,180}

\title{SciRAG: Adaptive, Citation-Aware, and Outline-Guided Retrieval and Synthesis for Scientific Literature}

\author{
\bf{
Hang Ding$^{\hspace{0.1em}\textcolor{SJTUred}{\boldsymbol{S}}}$~
Yilun Zhao$^{\hspace{0.1em}\textcolor{YaleBlue}{\boldsymbol{Y}}}$~
Tiansheng Hu$^{\hspace{0.1em}\textcolor{NYUPurple}{\boldsymbol{N}}}$~
Manasi Patwardhan$^{\hspace{0.1em}\textcolor{TataBlack}{\boldsymbol{T}}}$~
Arman Cohan$^{\hspace{0.1em}\textcolor{YaleBlue}{\boldsymbol{Y}}}$%
}
\\[6pt]
{
$^{\textcolor{SJTUred}{\boldsymbol{S}}}$Shanghai Jiao Tong University \quad
$^{\textcolor{YaleBlue}{\boldsymbol{Y}}}$Yale University \quad
$^{\textcolor{NYUPurple}{\boldsymbol{N}}}$NYU Shanghai \quad
$^{\textcolor{TataBlack}{\boldsymbol{T}}}$TCS Research
}\\[6pt]
\texttt{yilun.zhao@yale.edu}
}

\begin{document}

\maketitle

\begin{minipage}[t]{2\linewidth}
\vspace{-1.5cm}
  \centering
  \href{https://github.com/yale-nlp/SciRAG}{{\faGithub{}}\xspace\texttt{https://github.com/yale-nlp/SciRAG}} 
\vspace{0.5cm}
\end{minipage}

\begin{abstract}
The accelerating growth of scientific publications has intensified the need for scalable, trustworthy systems to synthesize knowledge across diverse literature. While recent retrieval-augmented generation (RAG) methods have improved access to scientific information, they often overlook citation graph structure, adapt poorly to complex queries, and yield fragmented, hard-to-verify syntheses. We introduce SciRAG, an open-source framework for scientific literature exploration that addresses these gaps through three key innovations: (1) adaptive retrieval that flexibly alternates between sequential and parallel evidence gathering; (2) citation-aware symbolic reasoning that leverages citation graphs to organize and filter supporting documents; and (3) outline-guided synthesis that plans, critiques, and refines answers to ensure coherence and transparent attribution. Extensive experiments across multiple benchmarks such as QASA and ScholarQA demonstrate that SciRAG outperforms prior systems in factual accuracy and synthesis quality, establishing a new foundation for reliable, large-scale scientific knowledge aggregation.
\end{abstract}

\section{Introduction}
With over four million journal articles published in 2024, scholarly output continues its decade-long 8\% annual growth~\citep{crossref2024}. The rise of preprint servers and open-access repositories has expanded scientific discourse, fostering cross-disciplinary discovery \citep{bornmann2015growth}, but also burdening researchers with reconciling fragmented findings, outpacing manual surveys and bibliometric tools \citep{beltagy2019scibertpretrainedlanguagemodel, asai2024openscholar, singh-etal-2024-scidqa}.

\begin{figure}[!t] 
    \centering
    \includegraphics[width=\linewidth]{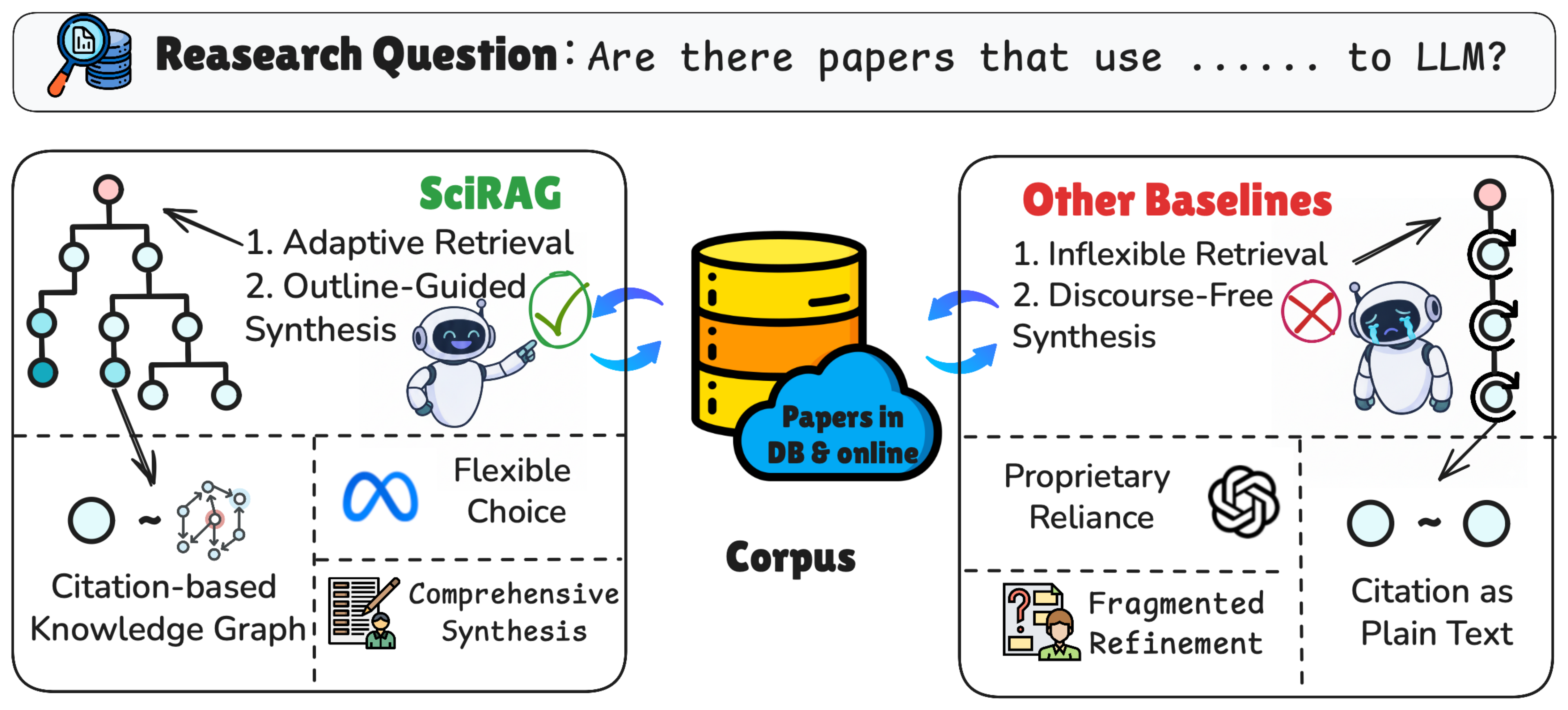} 
    \caption{An overview of SciRAG framework.}
    \label{fig:tea}
    \vskip -0.1in
\end{figure}

\emph{Retrieval-Augmented Generation} (RAG) has advanced rapidly since its introduction as a framework for knowledge-intensive NLP \citep{lewis2020retrieval}.
Recent systems couple LLMs with external search, significantly improving performance on knowledge-intensive benchmarks \citep{asai2024openscholar, zheng2024openresearcher, skarlinski2024languageagentsachievesuperhuman}.

When applied to scientific literature, however, current RAG systems still exhibit four key limitations: (1) \textbf{Superficial exploitation of citations}: references are treated as plain unstructured text rather than as structured relational entities, or, at best, single-hop backlinks, leaving the richer forward–backward citation graph
unused \citep{zhang2024litfm,agarwal2024litllm,bornmann2008citation}.
(2) \textbf{Inflexibility of retrieval}: queries are typically issued in a fixed, one-pass manner without adapting depth or coordinating orthogonal sub-topics \citep{asai2024openscholar,zheng2024openresearcher}. This limitation is especially pronounced for scientific literature, where complex multi-aspect questions (\eg combining theory, methodology, and application) demand dynamic and context-aware retrieval. 
(3) \textbf{Discourse-free synthesis}: models concatenate passage snippets without a global rhetorical plan, yielding answers that drift, overlook caveats, or conflate conflicting evidence \citep{skarlinski2024languageagentsachievesuperhuman}. 
(4) \textbf{Proprietary and costly frameworks}: many frameworks are proprietary and expensive, withholding models, indices, and workflows, which impedes reproducibility and further research. 

To address these gaps, we propose \textbf{SciRAG}, a retrieval and synthesis framework for scientific literature. SciRAG pioneers an adaptive architecture that dynamically integrates citation-driven reasoning over the literature graph, symbolic logic, and structured knowledge aggregation. To address the unique challenges of scientific literature QA, such as complex query structure, implicit cross-paper reasoning, and fragmented evidence, SciRAG is built around three tightly integrated components:

\begin{figure*}[htbp] 
    \centering
    \includegraphics[width=\linewidth]{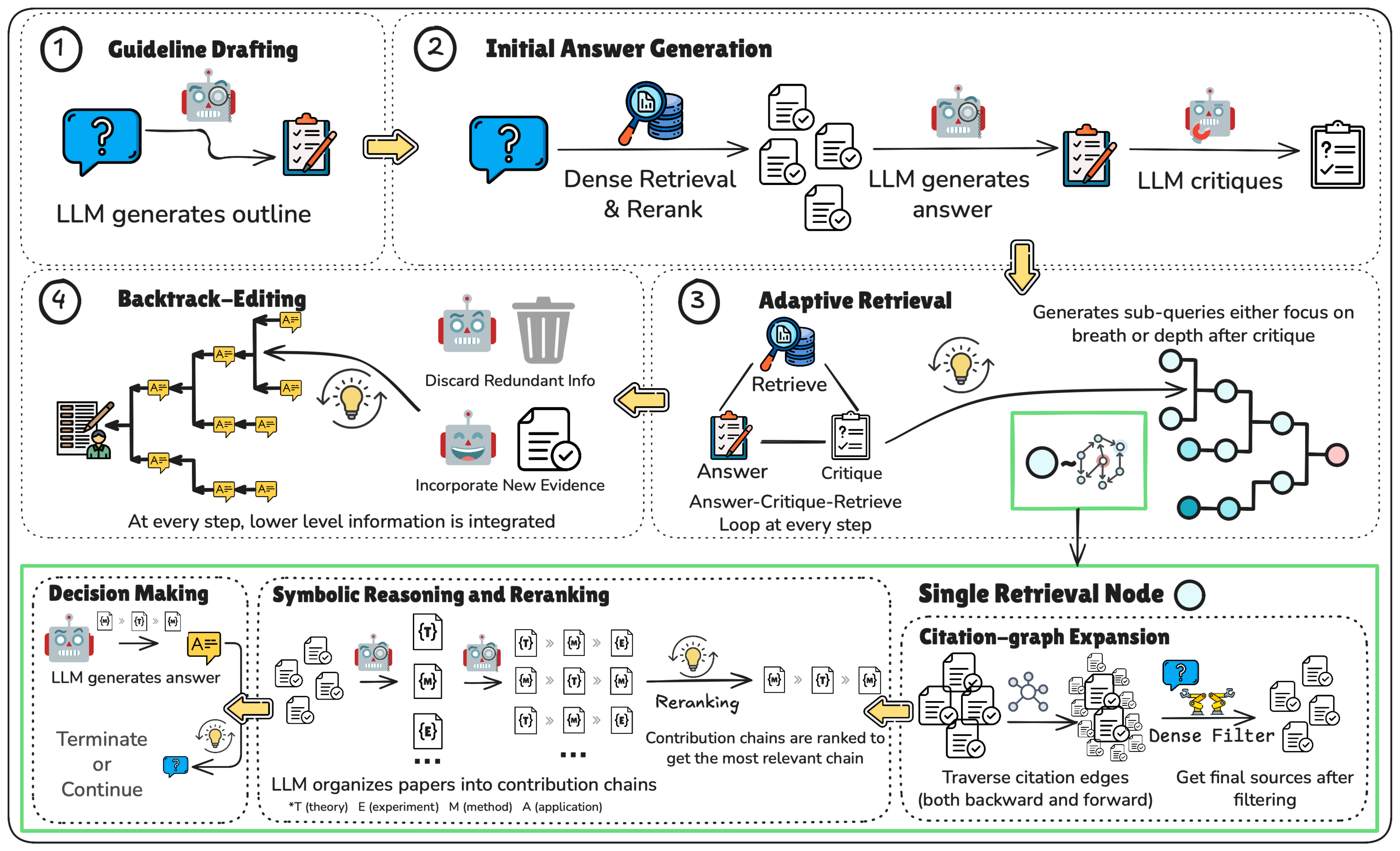}
    \caption{An illustration of the SciRAG pipeline. The process begins with guideline drafting and initial answer generation. Each retrieval node searches documents, decides whether to expand along the citation graph, builds contribution chains, and applies reasoning-based reranking to judge from current information whether to continue or stop. Adaptive retrieval integrates multiple nodes to balance sequential exploration for depth and parallel exploration for breadth. Finally, backtrack-editing consolidates all evidence and produces a coherent, well-documented answer.}
    \label{fig:over}
\end{figure*}

\begin{enumerate}[leftmargin=*, itemsep=0pt, topsep=0pt]
\item \textbf{Adaptive Retrieval.} A query-aware controller dynamically switches between sequential exploration, which is ideal for answering complex and in-depth questions, and parallel retrieval, which independently handles multiple sub-questions. This design enables flexible and comprehensive evidence gathering.
\item \textbf{Citation-Aware Symbolic Reasoning.} Explicitly traverses forward and backward citation paths in literature graphs to uncover conceptual relationships among studies, structuring retrieved evidence into interpretable contribution chains that guide both candidate reranking and logically grounded answer generation.
\item \textbf{Outline-Guided Synthesis.} SciRAG proposes an outline-guided synthesis module that first generates a coarse answer outline to structure retrieval, then iteratively identifies factual gaps, retrieves missing evidence, and refines the draft into a coherent and well-supported final answer.
\end{enumerate}

\vspace{3pt}
By integrating adaptive retrieval with citation centric symbolic reasoning, \textsc{SciRAG} establishes a new paradigm for trustworthy and scalable scientific knowledge synthesis. Extensive experiments on diverse open retrieval benchmarks, such as ScholarQA and PubMedQA, demonstrate that \textsc{SciRAG} consistently outperforms strong baselines including OpenScholar \citep{asai2024openscholar} and PaperQA2 \citep{skarlinski2024languageagentsachievesuperhuman}, achieving higher factual accuracy and overall relevance. Moreover, our analysis demonstrates the contributions of each component through ablation studies, verifies factual grounding via hallucination checks and case studies, and evaluates scalability under varying retrieval depths, confirming the framework’s robustness and interpretability.

\section{Related Work}

\paragraph{LLMs for Scientific Research.}
LLMs are beginning to automate a broad spectrum of scientific workflows, from idea generation and hypothesis formation \citep{baek2025researchagentiterativeresearchidea,yang2024largelanguagemodelsautomated} to code-level experiment design \citep{Huang2023MLAgentBenchEL,tian2024scicoderesearchcodingbenchmark} and literature-centric question answering \citep{asai2024openscholar,zheng2024openresearcher,skarlinski2024languageagentsachievesuperhuman}. Correspondingly, community benchmarks have evolved from single-paper fact checking ~\citep{wadden-etal-2020-fact} and abstractive QA~\citep{pmlr-v202-lee23n} to multi-disciplinary, multi-paper synthesis suites such as ScholarQABench \citep{asai2024openscholar,singh2025ai2scholarqaorganized}. 
These datasets highlight an emerging consensus: credible scientific assistance demands verifiable citations and wide coverage across disparate sub-fields. Such requirements are difficult for purely parametric LMs to satisfy, but we believe they can be effectively addressed by the integrated retrieval and reasoning design of \textsc{SciRAG}.

\paragraph{Graph-Enhanced RAG Systems.}
Several graph- and structure-aware RAG systems have been proposed to enhance open-domain QA by incorporating citation-graph propagation, section segmentation, or multi-hop retrieval. These systems include LitFM \citep{zhang2024litfm}, LitLLM \citep{agarwal2024litllm}, EfficientRAG \citep{zhuang2024efficientrag}, CoRAG \citep{wang2025chainofretrievalaugmentedgeneration}, DeepRAG \citep{guan2025deeprag}, and CG-RAG \citep{hu2025cgragresearchquestionanswering}. While these systems show promise on small-scale, static datasets for tasks such as title generation or local citation prediction, they often lack discourse-level reasoning and struggle with query-adaptive exploration over large open corpora. Relying on shallow, sequential chains, they fail to handle multi-hop knowledge trails, resulting in poor scalability and limited recall in open-domain synthesis across millions of documents, a core requirement in scientific literature synthesis. 

\paragraph{RAG Systems for Scientific Literature Tasks.}
To address citation fidelity, recent RAG systems combine LLMs with external corpora. OpenScholar \citep{asai2024openscholar} implements explicit citation verification and self-feedback; OpenResearcher \citep{zheng2024openresearcher} merges dense-sparse retrieval with adaptive query rewriting; and PaperQA2 \citep{skarlinski2024languageagentsachievesuperhuman} frames the task as a search-refine loop. However, these systems rely on sequential retrieval processes, which can overlook indirect evidence or lead to excessively large context windows when a query spans multiple theoretical threads. SciRAG, on the other hand, orchestrates parallel, citation-graph-aware retrieval, balancing high recall with precision and scaling efficiently to millions of papers. This approach enables SciRAG to address complex queries and multi-faceted tasks more effectively than its sequential counterparts.

\vspace{0.15in}
\section{\ours Framework}

Scientific literature retrieval and synthesis present distinct challenges beyond those in general-domain QA. Queries are often multi-faceted, requiring the integration of theoretical context, methodological details, and application-specific findings. Relevant evidence is scattered across papers connected by implicit conceptual links and complex citation structures, making isolated retrieval or naive summarization insufficient.

To this end, \textsc{SciRAG} introduces a tightly integrated pipeline comprising three complementary components: (1) \textbf{Outline-guided synthesis}, which structures answer generation through planning and iterative refinement; (2) \textbf{Citation-aware symbolic reasoning}, which constructs and prunes contribution chains through forward and backward citation expansion and symbolic reasoning; and (3) \textbf{Adaptive retrieval}, which dynamically alternates between sequential and parallel search based on query structure. Together, these modules address key obstacles in scientific QA, including retrieval inflexibility, reasoning opacity, and synthesis incoherence, while enabling transparent and verifiable responses grounded in literature. An overview of the full system is shown in Figure~\ref{fig:over}, and all the prompt templates of our proposed system can be found in Appendix~\ref{sec:prompt}.

\subsection{Outline-Guided Answer Aggregation with Reflective Refinement}

To overcome the fragmentation, inconsistency, and shallow structure of conventional retrieval-augmented methods, SciRAG employs an adaptive outline-guided aggregation procedure organized around a “plan–critic–solve” cycle. From the user’s query, it first derives a detailed outline that serves as a scaffold to keep retrieval and synthesis aligned with the intended scope and depth. Guided by this outline, SciRAG then enters a reflective refinement phase that critiques preliminary answers, diagnoses logical gaps, and triggers targeted retrieval to collect the additional evidence needed. Corrections are deferred to a backtracking edit phase, preserving coherence and reinforcing factual support.

Finally, SciRAG performs a bottom-up editing process: beginning from the deepest retrieval branches, each parent layer sequentially integrates the synthesized outputs of its children to iteratively revise and strengthen the draft. At each stage, newly retrieved evidence is incorporated, redundant or conflicting citations are pruned, and precise provenance links are maintained, producing coherent and well-grounded final answers.

\subsection{Citation-Graph Expansion with Symbolic Reasoning}
\label{subsec:cite-graph}

Typical RAG pipelines rely primarily on embedding similarity and thus overlook the \emph{structural} and \emph{logical} relations central to scientific discovery. To address this limitation, \textbf{SciRAG} enhances retrieval through two coordinated stages: (1) \emph{citation-graph expansion}, which broadens the search space under LLM guidance; and (2) \emph{symbolic reasoning}, which constructs, filters and ranks evidence by conceptual role. By combining these two stages, SciRAG produces a transparent, role-aware evidence set that supports robust multi-paper synthesis, going far beyond what similarity-only or shallow multi-hop systems can provide. For clarity, we provide a simplified snapshot in Appendix~\ref{sec:snap}, illustrating how contribution chains are constructed and how reranking and filtering are applied.  

\paragraph{Citation-graph expansion.} Starting from an initial embedding match set $P_0$, an LLM first judges whether $P_0$ suffices to answer query $q$. If not, SciRAG traverses both \emph{backward} and \emph{forward} edges (\(\leq 1\) hop) to assemble an enriched pool $P$. This step surfaces foundational works, key replications, and derivative applications that pure similarity search often misses, yielding a broader and more historically grounded candidate set for synthesis.  

\paragraph{Symbolic reasoning and reranking.} Each paper \(p \in P\), represented by its abstract, is segmented and tagged into conceptual roles: \textsf{T} (theory), \textsf{E} (experiment), \textsf{M} (method), \textsf{A} (application), etc., ensuring fine-grained understanding of its content. The LLM then analyzes these tagged segments to uncover conceptual links (\eg, “[1]\textsf{T}$\!\rightarrow$[2]\textsf{E}” when paper 1’s theory supports paper 2’s experiment), while pruning contradictory, tangential, or weakly supported branches. Pruning decisions reflect content relevance, query consistency, and logical coherence within the citation context. From the pruned graph, the model extracts \emph{contribution chains}, coherent multi-paper inference paths that collectively address the query. Rather than relying on ad hoc similarity or centrality scores, the LLM performs in-context reasoning to compare chains, evaluating their logical coherence, evidential completeness, and query relevance, then ranks the originating papers accordingly. Papers anchored in the strongest chains are promoted to the answer generator, while those with fragmented or unsupported reasoning are discarded, with justifications recorded for transparency.

\begin{algorithm}[!t]
\caption*{\small\textbf{Algorithms for Expansion \& Reasoning \& Reranking}}\label{alg:symbolic_modified}
\small
\begin{algorithmic}[1]
\Require query $q$, initial set $P_0$
\If{$\textsc{LLMJudge}(q, P_0)$}
\State $P \gets \textsc{ExpandGraph}(P_0)$
\Else
\State $P \gets P_0$
\EndIf
\ForAll{$p \in P$} \Comment{segment \& role tags}
\State $\text{seg}(p) \gets \textsc{Tag}(p)$
\EndFor
\State $G \gets \textsc{BuildRelationGraph}(P)$
\State $G \gets \textsc{PruneChains}(G)$ \Comment{remove contradictions}
\ForAll{$p \in P$}
\State $\textsc{Rank}(p) \gets \textsc{LLMReason}(q,p,G)$
\EndFor
\State \Return $\textsc{TopK}(P,\text{rank})$
\end{algorithmic}
\end{algorithm}

\begin{table*}[!t]
\centering
\small
\begin{tabular}{lllrl}
\toprule
\rowcolor{gray!25}
\textbf{\textit{Dataset}} & \textbf{\textit{Task}} & \textbf{\textit{Domain}} & \textbf{\textit{Size}} & \textbf{\textit{Metric}} \\
\midrule
SciFact      & Claim Verification & Biomedicine       & 208   & \Corr\ \& \Cite \\
PubMedQA     & Yes/No Judgement   & Biomedicine       & 843   & \Corr\ \& \Cite \\
QASA         & Q\&A               & Computer Science  & 1,375 & \Corr\ \& \Cite \\
\midrule
ScholarQA-\textsc{CS}    & Q\&A & Computer Science  & 100   & \Corr\ \& \Cite \\
ScholarQA-\textsc{Bio}   & Q\&A & Biomedicine       & 1,451 & \Cite \\
ScholarQA-\textsc{Neuro} & Q\&A & Neuroscience      & 1,308 & \Cite \\
ScholarQA-\textsc{Multi} & Q\&A & Multi-domain      & 108   & \Corr\ \& \Cite \\
\bottomrule
\end{tabular}
\caption{Overview of benchmarks evaluated in this study.  
Metrics: \Corr=Correctness Score,\ \Cite=Citation F1.}
\label{tab:benchs}
\end{table*}
\subsection{Adaptive Retrieval: Sequential and Parallel Mechanisms}

Traditional retrieval strategies for scientific literature often struggle to balance coverage with depth, especially for multifaceted queries. \textsc{SciRAG} addresses this limitation with an adaptive scheme that alternates between sequential and parallel retrieval, guided by the structural complexity and granularity of the user’s information need. To further boost precision, it retrieves short text fragments rather than entire documents and maps them back to sources for citation expansion. Passage-level retrieval not only raises topical relevance but also strengthens the foundation for evidence fusion. The entire process is driven by an “answer–critique–retrieval” loop, where each iteration diagnoses information gaps and generates sub-queries. Instead of being explicitly told to choose sequential or parallel search, the LLM automatically issues both deepening and broadening sub-queries, each launching its own retrieval thread and naturally combining depth- and breadth-oriented exploration.  

For queries requiring deep, context-dependent exploration, SciRAG builds retrieval rounds step by step: earlier results provide context that guides later searches, preserving logical continuity while progressively enriching the evidence. When a query decomposes into independent sub-questions, distinct retrieval threads operate concurrently, each targeting a specific sub-query (Appendix~\ref{sec:example}), thus improving efficiency and ensuring balanced coverage of divergent facets. By integrating sequential and parallel exploration with snippet-level retrieval, SciRAG achieves comprehensive, precise, and coherently organized coverage of the scientific literature, outperforming the rigid, one-dimensional pipelines of prior systems.  

\section{Experiment Setup}

\subsection{Baseline Systems}

To evaluate the effectiveness of SciRAG, we evaluate it against several strong baselines, briefly described below:
(1) \textbf{SciRAG}: Our proposed advanced framework. When using vector retrieval, We utilize the OpenScholar Datastore \cite{asai2024openscholar} as our vector retrieval corpus, which contains over 45 million papers and more than 200 million snippets. Throughout the entire pipeline, we use the standard GPT-4o Legacy model for retrieval, reasoning, and answer generation.
(2) \textbf{OpenScholar}~\cite{asai2024openscholar}: A large-scale scientific RAG system with an iterative self-feedback process to improve citation accuracy and content quality. It uses the same Datastore as SciRAG and provides four model versions (GPT-4o, Llama3.1-70B, OS-70B, OS-GPT4o), where OS-70B and OS-GPT4o are fine-tuned on scientific corpora.
(3)\textbf{PaperQA2}~\cite{skarlinski2024languageagentsachievesuperhuman}: A retrieval-augmented framework designed for literature synthesis. Implemented with its official open-source code and crawler, though our replication is limited by lack of access to private or license-protected papers.
(4) \textbf{GPT-4o with Online Search}~\cite{openai2024gpt4ocard}: GPT-4o augmented with real-time web search to enhance response accuracy.
(5) \textbf{Perplexity Pro}\footnote{\url{https://www.perplexity.ai/}}
: A commercial RAG system combining LLMs with real-time web search to deliver conversational responses with citations.

\begin{table*}[htbp]
\centering
\resizebox{\textwidth}{!}{
\begin{tabular}{lcc cc cc cc cc c c c}
\toprule
\multirow{2}{*}{\textbf{Method / Dataset}} 
& \multicolumn{2}{c}{\textbf{SciFact}} 
& \multicolumn{2}{c}{\textbf{PubMed}} 
& \multicolumn{2}{c}{\textbf{QASA}} 
& \multicolumn{2}{c}{\textbf{\textsc{CS}}} 
& \multicolumn{2}{c}{\textbf{\textsc{Multi}}}
& \textbf{\textsc{Bio}}
& \textbf{\textsc{Neuro}}
& \textbf{Cost}\\

\cmidrule(lr){2-3}
\cmidrule(lr){4-5}
\cmidrule(lr){6-7}
\cmidrule(lr){8-9}
\cmidrule(lr){10-11}
\cmidrule(lr){12-12}
\cmidrule(lr){13-13}
\cmidrule(lr){14-14}

& \Corr & \Cite
& \Corr & \Cite
& \Corr & \Cite
& \Corr & \Cite
& \Corr & \Cite
& \Cite
& \Cite
& \textbf{USD/query}\\

\midrule
\rowcolor{gray!30}\multicolumn{14}{c}{\textbf{\emph{Finetuned Baselines}}}\\
\midrule

OpenScholar-OS-70B    
& 82.1  &   47.5
&  79.6 &    74.0
&  \textbf{23.4} &    64.2
& 52.5 & 45.9
& 4.03 & 54.7  
& 55.9  
& 63.1 
&  0.01 \\

OpenScholar-OS-GPT4o    
& 81.3  &  56.5 
& 74.8  &   77.1
&  18.7 &   60.4
& 57.7 & 39.5
& 4.51 & 37.5 
& 51.5  
& 43.5  
&  0.12 \\
\midrule
\rowcolor{gray!30}\multicolumn{14}{c}{\textbf{\emph{Untuned / Legacy Baselines}}}\\
\midrule
GPT-4o        
& 77.8  &   0.0
& 65.8 &   0.0
&  21.2 &  0.0
& 45.0 &  0.1
& 4.01 &   0.7
&   0.2
&   0.1
& 0.06 \\

OpenScholar-GPT4o    
&  79.3  &   47.9
&  75.1  &   73.7
&  18.3 &   53.6
& 52.4 & 31.1
& 4.03 & 31.5  
& 36.3  
& 21.9  
&  0.12\\

OpenScholar-Llama3.1-70B    
& 78.2  &   42.5
&  77.4 &    71.1
&  22.7 &    63.6
& 48.5 & 24.5
& 4.24 & 41.4  
& 53.8  
& 58.1 
&  0.00 \\

PaperQA2\textsuperscript{\dag}       
&  -- &   --
& --  &  -- 
&  -- &   --
& 45.6 & 48.0
& 3.82 & 47.2  
& 56.7  
& 56.0  
& 0.3–2.3 \\

Perplexity Pro\textsuperscript{\dag}  
&  -- &   --
&  -- &   --
&  -- &   --
& 40.0 &   --
& 4.15 &   --
&   --
&   --
& 0.002 \\

\midrule
\rowcolor{gray!30}\multicolumn{14}{c}{\textbf{\emph{Ours (Untuned)}}}\\
\midrule
\rowcolor{cyan!6} \textbf{SciRAG(Llama3.1-70B)}
& 81.5 & 44.1
& 78.2  &   71.7
&  21.5 &   63.8
& 60.2 & 28.4
& 4.51 & 37.1
& 43.1
& 45.9
&  0.00\\

\rowcolor{cyan!6} \textbf{SciRAG(GPT-4O)}        
& \textbf{84.1} & 52.9
& \textbf{84.1}  &   74.8
&  19.2 &   54.2
& \textbf{69.0} & 34.0
& \textbf{4.79} & 37.8
& 44.8
& 36.2
&  0.16\\
\bottomrule
\end{tabular}
}
\vspace{1mm}
\caption{Performance comparison across multiple scientific QA datasets. Note that the evaluation scale for ScholarQA-\textsc{Multi} ranges from 0 to 5, whereas the other benchmarks adopt a 0–100 scale.\textsuperscript{\dag}: PaperQA2 is designed for multi-document synthesis and relies on a local PDF-based corpus, which is not publicly available. Perplexity Pro may incorporate non-scholarly sources and does not expose citation snippets, preventing full evaluation. As a result, we evaluate both baselines only on a subset of benchmarks.}
\label{tab:multi_dataset_comparison}
\end{table*}

\subsection{Evaluation Benchmarks}

For benchmark settings, we follow OpenScholar \cite{asai2024openscholar} and adopt the same four tasks: SciFact, PubMedQA, QASA, and ScholarQA, while explicitly adapting them to open-retrieval settings as detailed below. Together, these tasks cover diverse scientific domains and query types, enabling robust evaluation of answer synthesis.

\vspace{3pt}
\noindent\textbf{SciFact}~\citep{wadden-etal-2020-fact} is a biomedical claim verification benchmark, originally single-document. We adapt it to open retrieval by requiring evidence from a large corpus to verify claims as \textit{supported} or \textit{contradicted}, testing both fact-checking and retrieval.

\vspace{3pt}
\noindent\textbf{PubMedQA}~\citep{jin-etal-2019-pubmedqa} contains expert-written yes/no biomedical questions, originally paired with abstracts. We convert it into open retrieval, where models must locate literature to determine the binary answer.

\vspace{3pt}
\noindent\textbf{QASA}~\citep{pmlr-v202-lee23n} contains reasoning-heavy questions from single AI/ML papers. We adapt it to open retrieval, where models must locate source documents before answering.

\vspace{3pt}
\noindent\textbf{ScholarQA}~\citep{asai2024openscholar} evaluates multi-document synthesis for literature review questions across four domains: computer science (\textsc{CS}), biomedicine (\textsc{Bio}), neuroscience (\textsc{Neuro}), and a mixed set (\textsc{Multi}). \textsc{CS} includes expert references and rubrics; \textsc{Bio} and \textsc{Neuro} provide curated questions demanding deep synthesis; \textsc{Multi} offers long-form answers with citations for detailed assessment of coverage and citation quality.

Detailed dataset statistics are shown in Table~\ref{tab:benchs}.

\subsection{Evaluation Protocols}
\paragraph{Automated Evaluation.}
We adopt two core metrics. \textbf{Correctness Score} evaluates factual consistency and relevance. For SciFact and PubMedQA, we use Exact Match against expert-labeled binary answers. QASA is scored by ROUGE-L overlap with references. ScholarQA-\textsc{CS} uses expert rubrics specifying must-have and nice-to-have elements, while ScholarQA-\textsc{Multi} is assessed via \textit{Prometheus-8x7b-v2.0}~\cite{kim-etal-2024-prometheus}, which rates relevance, coverage, and organization. \textbf{Citation F1} captures citation accuracy as the harmonic mean of precision (citation supports claim) and recall (all citation-worthy claims are cited). It is used for all datasets and serves as the sole metric for ScholarQA-\textsc{Bio} and \textsc{Neuro}. These metrics support a realistic and comprehensive evaluation of SciRAG in open-retrieval scientific QA.

\paragraph{Human Evaluation.} \label{para:human}
We conducted a human evaluation with three expert annotators, each holding at least a Master's degree in CS. The evaluation compared answers generated by \ours, OpenScholar, and PaperQA2 for 30 randomly sampled queries from the ScholarQA-\textsc{CS} dataset. Each annotator evaluated the three answers for each query, scoring them on four aspects: \emph{Relevance}, \emph{Coverage}, \emph{Organization}, and \emph{Overall Usefulness}, using a 1-5 scale. The final score for each aspect was calculated as the average of the three annotators' scores. The evaluation criteria are further detailed in Appendix~\ref{app:human-eval-criteria}.

\section{Experiment}
\label{sec:results}

We perform extensive experiments to evaluate the effectiveness, reliability, and reasoning capabilities of SciRAG across multiple scientific QA datasets.

\subsection{Main Results}

As shown in Table ~\ref{tab:multi_dataset_comparison}, across all evaluated benchmarks SciRAG consistently delivers top-tier answer quality, ranking first in correctness score on 4 out of 5 datasets. This demonstrates the effectiveness of our symbolic reasoning and outline-guided synthesis in producing coherent, accurate, and logically structured responses. Notably, SciRAG outperforms strong baselines such as OpenScholar-OS-GPT4o and PaperQA2, with clear gains on SciFact (+2.8), PubMedQA (+9.3), ScholarQA-\textsc{CS} (+11.3), and ScholarQA-\textsc{Multi} (+0.28) compared to OpenScholar-OS-GPT4o, highlighting its strength in tackling complex, multi-faceted queries across diverse scientific domains.

\begin{figure*}[!t]
  \centering
  \includegraphics[width=0.9\textwidth]{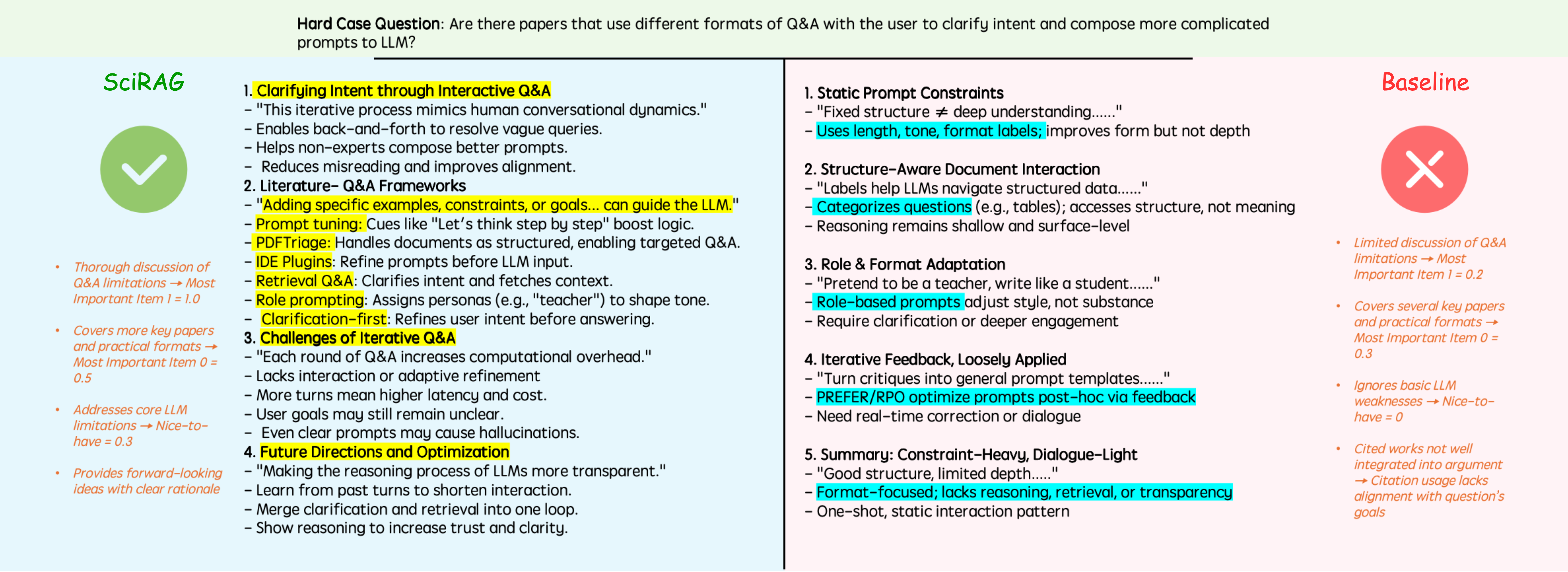}
  \caption{An example comparing \ours with a representative baseline.}
  \label{fig:example}
  \vskip -0.05in
\end{figure*}

However, SciRAG does not achieve the highest citation F1 on several benchmarks. This gap stems primarily from two factors. First, our system uses the general GPT-4o model, which lacks the specialized fine-tuning for citation accuracy that OS-GPT4o benefits from. Second, SciRAG sometimes generates summary-style or inferential statements that, while logically sound, do not have explicit one-to-one citation links, leading to lower scores under strict F1 evaluation criteria. Nevertheless, SciRAG maintains competitive citation precision scores, supported by robust citation graph expansion, symbolic reranking, and citation verification modules. Further manual analysis reveals that many seemingly uncited statements are in fact implicitly supported through multi-hop citation chains, a nuance missed by surface-level metrics.

\subsection{Human Evaluation}

\begin{table}[!t]
\centering
\small
\renewcommand{\arraystretch}{1.08}
\begin{tabular}{@{}lccc@{}}
\toprule
\textbf{Metric} & \textbf{SciRAG} & \textbf{OpenScholar} & \textbf{PaperQA2} \\
\midrule
Org. & \textbf{3.83} & 3.75 {\scriptsize\textcolor{red}{(-0.08)}} & 3.69 {\scriptsize\textcolor{red}{(-0.14)}} \\
Cov.     & \textbf{4.00} & 3.51 {\scriptsize\textcolor{red}{(-0.49)}} & 2.62 {\scriptsize\textcolor{red}{(-1.38)}} \\
Rel.    & 3.49          & \textbf{3.63} {\scriptsize\textcolor{blue}{(+0.14)}} & 3.38 {\scriptsize\textcolor{red}{(-0.11)}} \\
Usef.   & \textbf{3.67} & 3.30 {\scriptsize\textcolor{red}{(-0.37)}} & 2.72 {\scriptsize\textcolor{red}{(-0.95)}} \\
\hdashline
Avg.      & \textbf{3.75} & 3.55 {\scriptsize\textcolor{red}{(-0.20)}} & 3.10 {\scriptsize\textcolor{red}{(-0.65)}} \\
\bottomrule
\end{tabular}
\caption{Human evaluation on ScholarQA-\textsc{CS}. Numbers in parentheses show the difference vs.\ SciRAG.}
\label{tab:human-eval-result}
\end{table}

We also conducted a qualitative evaluation of the generated outputs using three expert annotators The protocol is detailed in Section \ref{para:human}.               
The evaluation result is shown in Figure ~\ref{tab:human-eval-result}. Among the four aspects, \ours leads in \emph{Organization}, \emph{Coverage}, \emph{Usefulness}, demonstrating \ours could generate a well-organized, comprehensive and useful result to scientific literature queries. \ours scores lower than OpenScholar in \emph{Relevance}, primarily because its answers often include additional background or contextual information. While this content is designed to support comprehensive understanding, evaluators may sometimes perceive such introductory material as unrelated to the main question. To further assess reliability, we examined inter-annotator agreement. The average pairwise correlation between annotator scores was consistently high (around 0.87), confirming that the evaluations are stable and not driven by individual annotator variance.

\subsection{Hallucination Analysis}
To assess the reliability of uncited statements, we conducted a sentence-level evaluation on 100 randomly sampled generation cases, using the same granularity as citation F1. In each case, all sentences without explicit citations were extracted and reviewed by three independent LLM judges (GPT-4o, DeepSeek-R1, and Gemini 2.5 Pro). The judges determined whether each sentence was supported by the retrieved context, either directly or through multi-hop reasoning. The proportion of sentences judged unsupported was 6.0\% for GPT-4o, 5.3\% for DeepSeek-R1, and 6.6\% for Gemini 2.5 Pro. These highly consistent results across models indicate that the vast majority of SciRAG’s responses remain well-grounded in evidence.

\subsection{Case Studies}
\label{sec:case-study}

To further evaluate the performance of \ours against the baseline, we conduct a representative hard case study, with the LLM-based evaluation results visualized in Figure~\ref{fig:example}.
Compared to the baseline, which offers only a limited discussion of Q\&A limitations, partial coverage of related works, and poorly aligned citations, \ours provides a more thorough treatment of Q\&A challenges, integrates a wider range of key papers and practical formats, and explicitly addresses LLM weaknesses while offering forward-looking insights. This contrast shows that \ours produces answers that are both more comprehensive and analytically grounded. For completeness, we also include in Table~\ref{tab:user-study} in Appendix the human evaluation of this case, where expert annotators likewise preferred \ours for its broader coverage and deeper reasoning.

\subsection{Ablation Studies on Reasoning Modules}
\label{sec:ablation}

We disentangle the contributions of SciRAG’s two core reasoning modules—\emph{symbolic reranking} and the \emph{outline-based meta-planner}—through controlled ablations on three representative benchmarks (Table~\ref{tab:ablation}). Removing either module leads to performance degradation, but in different ways. 

\textbf{Symbolic reranking} governs evidence precision: replacing it with a dense reranker results in sharp drops of 10.7 on SciFact, 10.4 on ScholarQA-\textsc{CS}, and 0.34 on ScholarQA-\textsc{Multi}. Dense models rely only on embedding similarity, often retrieving papers with overlapping terms but divergent focus. In contrast, our symbolic reranker filters evidence by conceptual role and logical consistency within citation chains, while also providing reasoning traces and rationales that enhance transparency.

The effects of retrieval strategy are similarly clear. \textbf{Sequential-only} and \textbf{Parallel-only} variants show that depth preserves logical continuity and breadth expands coverage, yet both fall short of the adaptive integration that combines their strengths. \textbf{The meta-planner}, meanwhile, drives global answer structure. Without it, the model loses rhetorical guidance, producing fragmented or incoherent responses. This limitation is most pronounced in complex synthesis tasks, where performance drops by 16.9 points on ScholarQA-\textsc{CS} and 0.72 on ScholarQA-\textsc{Multi}, underscoring the planner’s role in coordinating multi-branch integration.

Taken together, symbolic reranking and the meta-planner form the backbone of SciRAG’s reasoning pipeline: the former sharpens evidence selection, while the latter orchestrates synthesis. Their complementary roles are essential for producing precise, well-structured, and verifiable answers.

\begin{table}[!t]
\centering
\small
\begin{tabular}{@{}p{2.4cm}lll@{}}  
\toprule
\textbf{Model} & \textbf{SciFact} & \textbf{ScQA-CS} & \textbf{ScQA-Multi} \\
\midrule
Full SciRAG & \textbf{84.1} & \textbf{69.0} & \textbf{4.79} \\
\addlinespace[3pt]
\hdashline
\addlinespace[3pt]
\quad \emph{w.} Dense Rerank & 73.4 {\scriptsize\textcolor{red}{(-10.7)}} & 
58.6 {\scriptsize\textcolor{red}{(-10.4)}} & 
4.45 {\scriptsize\textcolor{red}{(-0.34)}} \\
\quad Sequential-only    & 75.7 {\scriptsize\textcolor{red}{(-8.4)}} & 
59.1 {\scriptsize\textcolor{red}{(-9.9)}} & 
4.41 {\scriptsize\textcolor{red}{(-0.38)}} \\
\quad Parallel-only      & 77.3 {\scriptsize\textcolor{red}{(-6.8)}} & 
58.2 {\scriptsize\textcolor{red}{(-10.8)}} & 
4.56 {\scriptsize\textcolor{red}{(-0.23)}} \\
\addlinespace[3pt]
\hdashline
\addlinespace[3pt]
\quad w/o Planner & 76.1 {\scriptsize\textcolor{red}{(-8.0)}} & 
52.1 {\scriptsize\textcolor{red}{(-16.9)}} & 
4.07 {\scriptsize\textcolor{red}{(-0.72)}} \\
\bottomrule
\end{tabular}
\caption{Impact of removing each module of \ours.}
\label{tab:ablation}
\end{table}

\subsection{Scaling Behavior: Retrieval Tree Depth}
\label{sec:scaling}

\begin{figure}[h]
    \centering
    \begin{minipage}{0.48\linewidth}
        \centering
        \includegraphics[width=\linewidth]{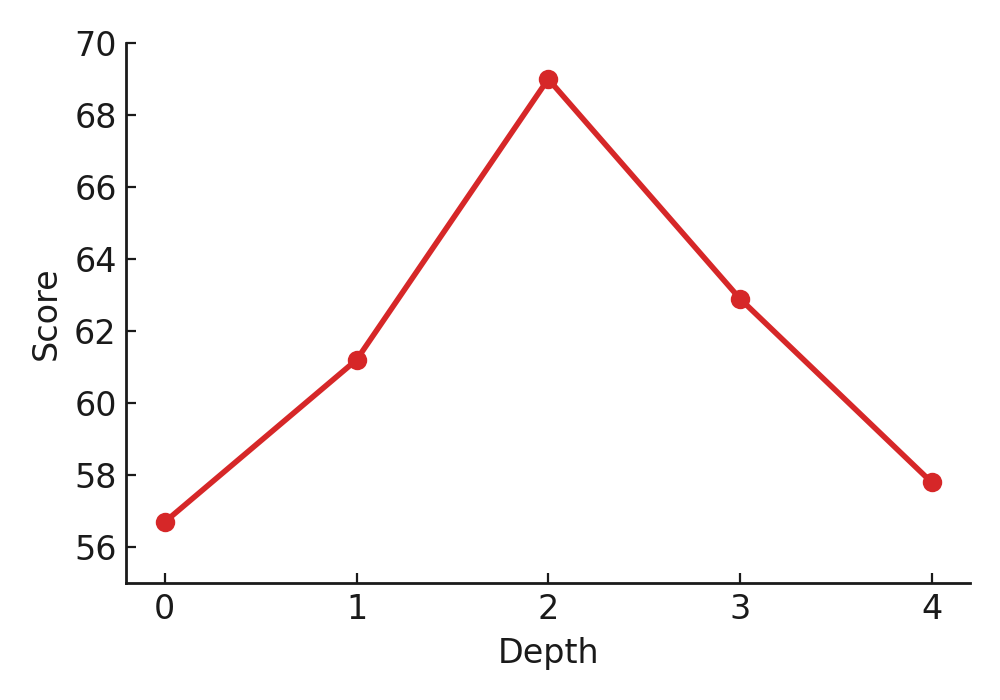}
        \caption*{(d) ScholarQA-\textsc{CS}}
    \end{minipage}
    \hfill
    \begin{minipage}{0.48\linewidth}
        \centering
        \includegraphics[width=\linewidth]{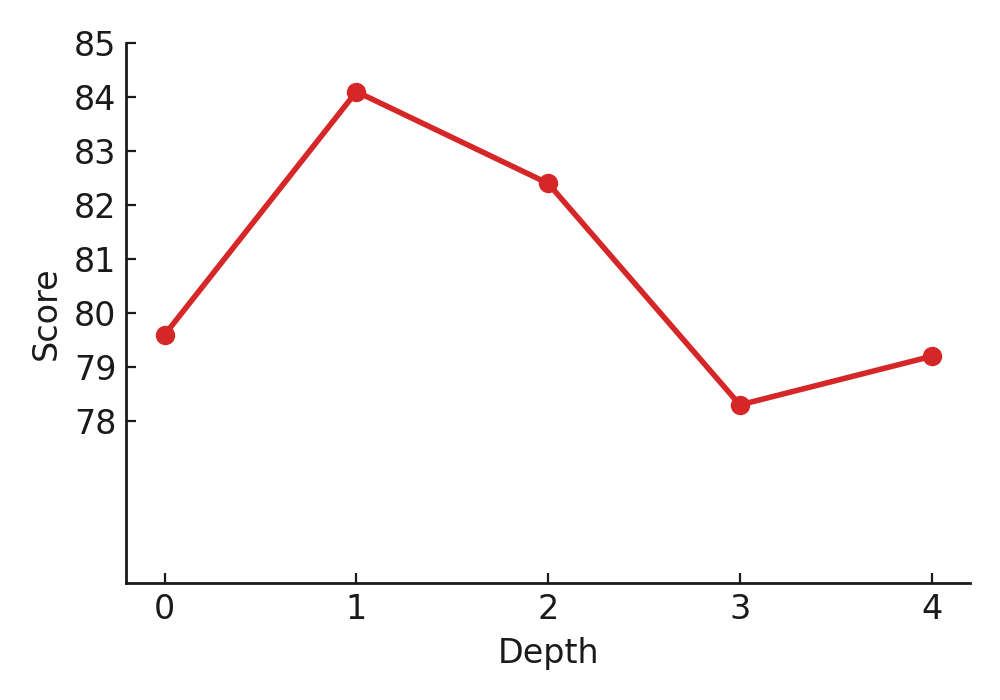}
        \caption*{(b) SciFact}
    \end{minipage}
    
    \vspace{0.5em}
    
    \begin{minipage}{0.48\linewidth}
        \centering
        \includegraphics[width=\linewidth]{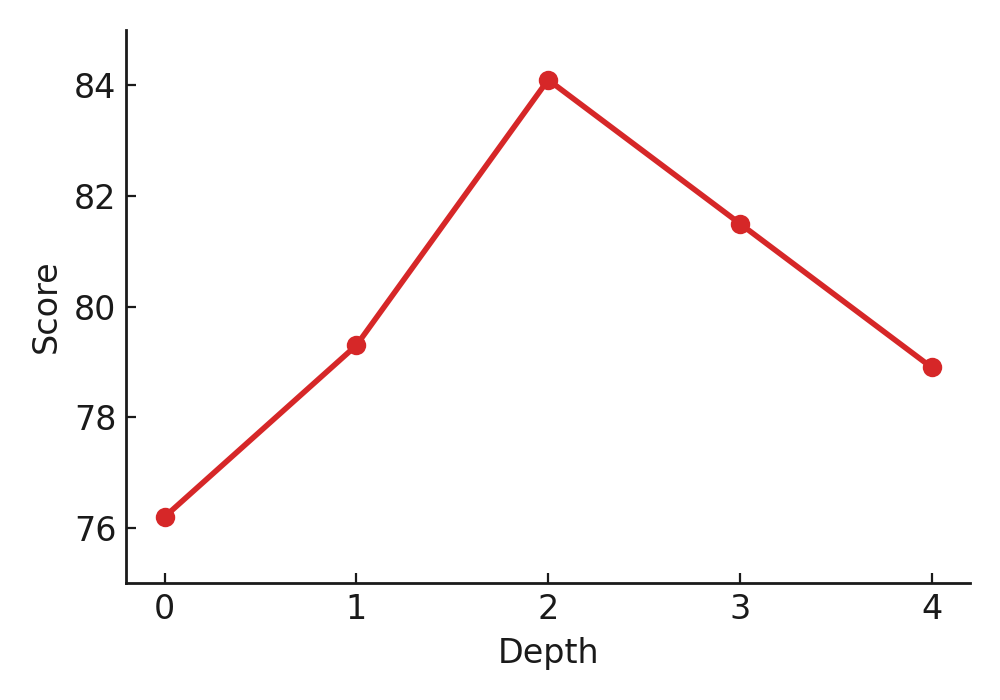}
        \caption*{(c) PubMedQA}
    \end{minipage}
    \hfill
    \begin{minipage}{0.48\linewidth}
        \centering
        \includegraphics[width=\linewidth]{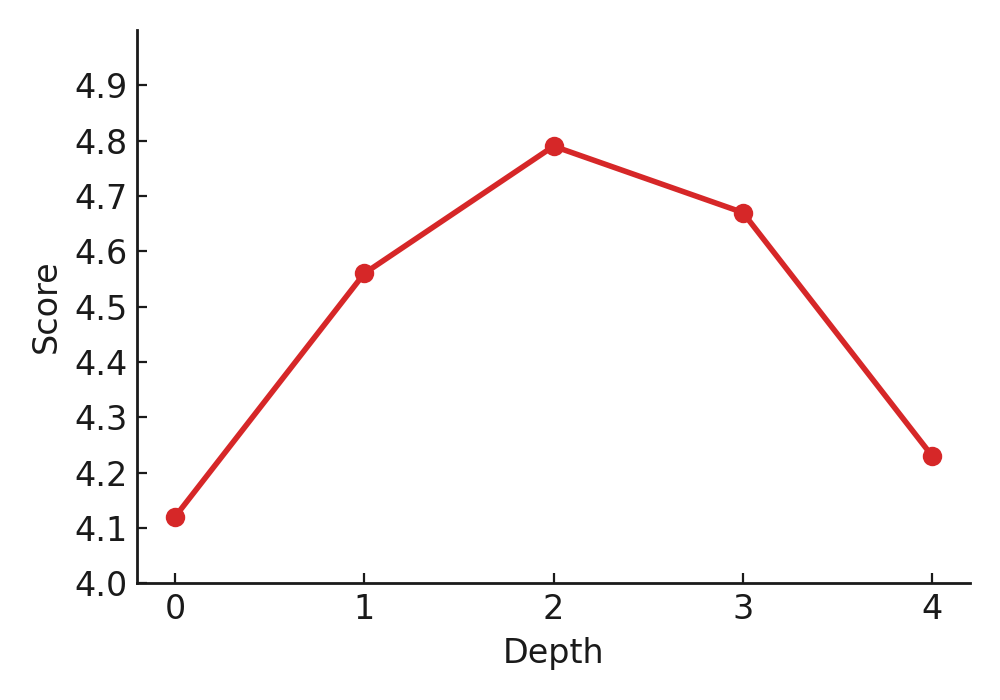}
        \caption*{(a) ScholarQA-\textsc{Multi}}
    \end{minipage}
    \caption{Effect of retrieval depth on answer quality across four benchmarks.}
    \label{fig:depth_scaling_all}
\end{figure}

We further evaluate the robustness of SciRAG under increasing retrieval tree depths. Figure~\ref{fig:depth_scaling_all} illustrates answer quality trends across ScholarQA-\textsc{CS}, SciFact, PubMedQA, and ScholarQA-\textsc{Multi} benchmarks as retrieval depth varies from 0 to 4 levels.
A common trend emerges across all benchmarks: answer quality improves as retrieval depth increases from 0 to 2, before deteriorating at greater depths. Performance consistently peaks at depth~2 (or 1), suggesting an optimal balance between sufficient context and avoiding noise. For instance, PubMedQA improves from 76.2 (depth 0) to 84.1 (depth 2), indicating successful integration of relevant evidence. However, further expansion (depth 3–4) reduces quality, likely due to citation noise and tangential content overwhelming the model’s reasoning capacity.
This analysis confirms our design choice of using moderate depths, maximizing supportive evidence while minimizing irrelevant information. Consequently, these findings not only provide empirical validation for SciRAG’s adaptive retrieval mechanism but also inform practical guidelines for optimal retrieval-depth selection in future research.
\section{Conclusion}

We presented \textbf{SciRAG}, a novel retrieval-augmented generation framework designed specifically for scientific literature exploration. By integrating adaptive retrieval, citation-aware symbolic reasoning, and outline-guided synthesis, SciRAG addresses key limitations of existing approaches, including inflexible retrieval, superficial citation usage, and fragmented answer construction. Comprehensive experiments on open-retrieval benchmarks such as SciFact, PubMedQA, and ScholarQA demonstrate that SciRAG achieves state-of-the-art performance in factual accuracy, coherence, and overall usefulness. It consistently generates coherent, well-organized responses grounded in verifiable evidence, as further validated by expert human evaluations.Looking forward, SciRAG provides a scalable and transparent foundation for trustworthy scientific question answering, with strong potential to assist researchers in navigating the ever-growing volume and complexity of scholarly literature.
\section*{Limitations} 
Despite its strong performance, \textbf{SciRAG} has certain limitations. First, it relies on general-purpose language models such as GPT-4o and Llama-3.1, which are not fine-tuned for scientific citation accuracy and may miss precise attribution. Second, the symbolic reasoning and outline-based synthesis introduce non-trivial computational overhead, which may affect real-time applicability in large-scale deployments. Also, the human evaluation was conducted with a limited number of expert annotators, which may not capture disciplinary variance. Future work could explore domain-specific model tuning and lightweight alternatives to improve both precision and efficiency.

\bibliography{custom}

\appendix
\onecolumn
\section{Case Study}

\begin{table}[H]
\centering
\resizebox{\textwidth}{!}{
\begin{tabular}{|p{0.95\textwidth}|}
\hline
\textbf{Hard Case Question:} Are there papers that use different formats of Q\&A with the user to clarify intent and compose more complicated prompts to LLM?
 \\
\hline
\textbf{SciRAG:} 1. Significance of Using Q\&A Formats for Intent Clarification and Enhanced Prompt Composition in LLMs

The use of Q\&A formats to clarify user intent and enhance prompt composition is gaining prominence as an effective strategy to improve the quality [3][6] and relevance of interactions ...

2. Survey of Literature on Q\&A-Based Frameworks for Refining User Intent and Generating Complex Prompts

Recent studies have investigated various Q\&A-based frameworks and methodologies for improving user interaction with LLMs[4]. These frameworks emphasize iterative clarification, intent disambiguation, and the construction of structured, context-rich prompts.
One notable approach is ......

3. Challenges and Limitations of Q\&A Formats for Intent Clarification in LLMs

Despite their advantages, Q\&A-based frameworks for intent clarification face several challenges and limitations. One significant issue is scalability, particularly when .....[5]

4.Conclusion and Future Directions

In summary, Q\&A-based frameworks offer a promising approach to improving the quality and relevance of interactions with LLMs by refining user intent and ...[10]
\\
\textbf{Fine-grained Assessments:} \fbox{Org} 5.0 \fbox{Rel} 5.0 \fbox{Con} 5.0 \\
\hline
\textbf{OS-GPT4o} Yes, several papers explore the use of different formats for question-and-answer interactions to clarify user intent and construct more sophisticated prompts \ldots[2]

Structured and Granular Prompt Engineering

Structured and granular prompt engineering involves designing prompts that align with specific structural constraints, such as numerical, descriptive, and formatting requirements \ldots[0]

Dynamic Interaction with Structured Documents

This section discusses user interactions with structured documents, focusing on numerical, descriptive, and formatting structural constraints. \ldots[2]

Adaptive and Role-Specific Prompts

Taxonomy Numerical Constraining text to a set or a bounded number of words, sentences, or paragraphs is valuable in all aspects of writing  \ldots[2]

Iterative Feedback and Post-Processing

Iterative feedback and post-processing are valuable in improving the quality of responses generated by language models. Specifically, incorporating user feedback into \ldots[1]

Summary and Implications

Collectively, these studies highlight the importance of using structured, adaptive, and iterative Q\&A formats to clarify user intent and enhance prompt engineering for LLMs. Techniques such as feedback loops \ldots[3]
\\
\textbf{Fine-grained Assessments:} \fbox{Org} 4.0 \fbox{Rel} 5.0 \fbox{Con} 4.0\\
\hline
\textbf{Preference:} SciRAG\\
\textbf{Explanation:} Compared to the baseline, SciRAG offers more comprehensive topic coverage and a clearer structural flow. It systematically introduces key concepts, explores multiple relevant methods, and highlights potential limitations, enabling a deeper understanding of the question. In contrast, the baseline answer is fragmented, with less clear organization and limited depth across sections, making it harder to grasp the full scope of the discussion. \\
\hline
\end{tabular}
}
\caption{Evaluation Case Study}
\label{tab:user-study}
\end{table}

\FloatBarrier

\section{Human Evaluation Criteria}\label{app:human-eval-criteria}
Inspired by OPENSCHOLAR \citep{asai2024openscholar}, we define four fine-grained aspects to evaluate the generation. We ask the evaluator to give a score from 1 to 5 with regarding to the four aspects, \emph{Relevance, Coverage, Organization and Overall Usefulness}. The definition and instruction of scoring is shown in table \ref{tab:criteria}
\begin{table*}[htbp]
\centering
\resizebox{0.8\textwidth}{!}{
\begin{tabular}{>{\centering\arraybackslash}m{2.5cm}|
                >{\centering\arraybackslash}m{6cm}|
                >{\centering\arraybackslash}m{6cm}}
\hline
\textbf{Aspect} & \textbf{Definition} & \textbf{Instructions} \\
\hline
Organization & Evaluate if the output is well-organized and logically structured. & Score 1 means the response is disorganized, with no clear structure. \newline Score 5 means the response is exceptionally well-organized, with a flawless logical structure \\
\hline
Coverage & Evaluate if the output provides sufficient coverage and amount of information. & Score 1 means the answer misses most of the key areas with few resources. \newline Score 5 means the answer covers a diverse range of papers and viewpoints offering a thorough overview of the area. \\
\hline
Relevance & Evaluate if the response stay on topic and maintain a clear focus to provide a useful response to the question. & Score 1 means the content significantly deviates from the original question. \newline Score 5 means the response remains tightly centered on the subject matter with enough depth for every piece of information. \\
\hline
Overall Usefulness & Evaluate if the output contains useful information to fulfill the information needs. & Score 1 means the response does not answer the question or provides rather confusing information. \newline Score 5 means the response provides a comprehensive overview of the area, and sufficiently answers the question without additional references.\\
\hline
\end{tabular}
}
\caption{Evaluation Criteria Descriptions}
\label{tab:criteria}
\end{table*}

\section{Case Examples}\label{sec:example}
Below is an example of our query depth and parallel dynamic retrieval. In this example, the sub-queries 1 and 3 were further refined through deeper retrieval after initial searching to gather more detailed information. On the other hand, sub-queries 2 and 4 employed a parallel retrieval strategy, where multiple sub-queries were independently expanded at the same time, thereby broadening the scope of retrieval and ensuring comprehensive coverage of multidimensional information.

\newtcolorbox[auto counter, number within=section]{mybox}[2][]{colframe=blue!20!white, colback=blue!5, coltitle=black, fonttitle=\bfseries, title=#2,#1}

\begin{tcolorbox}[colframe=red!30!white, colback=red!5, coltitle=black, fonttitle=\bfseries, title=Root Question]
    Are there papers that use different formats of Q\&A with the user to clarify intent and compose more complicated prompts to LLM?
\end{tcolorbox}

\begin{mybox}[label=sec1]{1. What are the different clarification question formats used to clarify user intent in Q\&A systems with LLMs?}
    \begin{tcolorbox}[colframe=green!30!white, colback=green!5, coltitle=black, fonttitle=\bfseries, title=1.1]
        What are the different clarification question formats used to clarify user intent in Q\&A systems with LLMs?
    \end{tcolorbox}
\end{mybox}

\begin{mybox}[label=sec2]{2. What performance metrics are used to assess the impact of clarification and multi-step questioning techniques on LLM accuracy?}
    \begin{tcolorbox}[colframe=orange!30!white, colback=orange!5, coltitle=black, fonttitle=\bfseries, title=2.1]
        What are the most widely used benchmarks to assess LLM performance in tasks involving clarification and multi-step questioning?
    \end{tcolorbox}
    \begin{tcolorbox}[colframe=purple!30!white, colback=purple!5, coltitle=black, fonttitle=\bfseries, title=2.2]
        How is user engagement specifically measured when applying clarification or multi-step questioning techniques in LLM-based systems?
    \end{tcolorbox}
\end{mybox}

\begin{mybox}[label=sec3]{3. How does multi-step question generation work in LLMs to improve the precision of prompts?}
    \begin{tcolorbox}[colframe=yellow!30!white, colback=yellow!5, coltitle=black, fonttitle=\bfseries, title=3.1]
        What techniques are used for the adaptive generation of multi-step questions in LLMs to ensure prompt precision?
    \end{tcolorbox}
\end{mybox}

\begin{mybox}[label=sec4]{4. What are some real-world applications where multi-step Q\&A strategies have been successfully applied to LLM systems?}
    \begin{tcolorbox}[colframe=cyan!30!white, colback=cyan!5, coltitle=black, fonttitle=\bfseries, title=4.1]
        What measurable impact has multi-step Q\&A had on customer satisfaction in service-based industries?
    \end{tcolorbox}
    \begin{tcolorbox}[colframe=violet!30!white, colback=violet!5, coltitle=black, fonttitle=\bfseries, title=4.2]
        How have multi-step Q\&A strategies been specifically applied in healthcare for improving diagnostic accuracy or patient interactions?
    \end{tcolorbox}
\end{mybox}

\section{Simplified Snapshot of Contribution Chains and Symbolic Reranking}\label{sec:snap}
To complement the method description in §3.2 , we provide a compact snapshot that makes the mechanics of contribution-chain construction and symbolic reranking/filtering more transparent. 

\begin{jsonbox}
{ "query": "How do different formats of multi-turn Q&A (e.g., structured, iterative, and conversational) specifically address user intent clarification and prompt accuracy in complex scenarios like legal, medical, or technical domains?", 

"path": "Are there papers that use different formats of Q&A with the user to clarify intent and compose more complicated prompts to LLM? -> How do different formats of multi-turn Q&A (e.g., structured, iterative, and conversational) specifically address user intent clarification and prompt accuracy in complex scenarios like legal, medical, or technical domains?",

"papers_count": 10,

"step1_analysis": { "output": "{"papers": [{"paper_index": 1,"segments": [{"label": "T","description": "Explores clarifying questions as a key property in conversational search and mixed-initiative IR, highlighting the role of multi-turn Q&A.","relevance": "High"},{"label": "E","description": "Compares GPT-based simulators with human query reformulation patterns and discusses evaluation challenges for multi-turn clarification.","relevance": "Medium"}]},........]}" },

"step2_relationships": { "output": "{"relationships": [{"symbol": "[1]T -> [Q]","rationale": "Paper 1's exploration of clarifying questions in conversational search informs the understanding of multi-turn Q&A formats for user intent clarification."},{"symbol": "[3]T -> [5]M","rationale": "Paper 3's discussion of multi-turn LLM performance challenges addresses issues relevant to improving accuracy in complex scenarios from the problems coming from Paper 5's methodology."},........]}" },

"step3_analysis": { Crafting analysis plan > We need to follow the instructions step by step. I should avoid web-based responses, focusing on a detailed yet concise breakdown. I'll identify key papers, highlight contradictions and irrelevancies, and stick to the exact data provided. I'll include symbolic relationships and ensure responses address the query directly, using numbered lists for clarity. No extra claims should be made.> Breaking down paper contributions > I went through the symbolic relationships and the paper breakdowns and produced a focused, step-by-step analysis of how the set of papers coheres (or not) with the retrieval query: “How do different formats of multi-turn Q&A (structured, iterative, conversational) specifically address user intent clarification and prompt accuracy in complex scenarios (legal, medical, technical)?” # 1) Core papers (most central) and why they matter These papers most directly speak to mechanisms, failure modes, and empirical tests of multi-turn Q&A styles that aim to clarify intent and improve prompt accuracy. 1. Paper 1 — (T + E) - T (Theory): Directly frames clarifying questions as core to conversational search and mixed-initiative interaction. This is central because clarifying Qs are the foundational behavior across structured, iterative, and conversational formats. - E (Evaluation): ......

Simulator realism / external validity tension - Papers involved: Paper 1 (E) vs Paper 5 (M/E). - Tension: Paper 1’s evaluation of GPT-based simulators diverging from human query reformulations raises questions about how well simulated user studies (including USi in Paper 5) generalize to real users. Paper 5 claims accurate simulation via crowdsourced validation, but Paper 1 flags evaluation challenges. - Implication: Any conclusions about format effectiveness based mainly on simulated interactions need corroboration with real users, or at least strong validation against human reformulations. 2. Performance drop (Paper 3) vs targeted improvements (Papers 6 & 8) - .......

Mapping the papers to the three format families (quick cheat-map) - Structured Q&A (explicit templates, decision policies, feature-driven): Paper 2 (E), Paper 6 (T/E), Paper 3 (T). - Iterative Q&A (stepwise clarification, short loops): Paper 1 (T), Paper 5 (M/E), Paper 3 (E). - Conversational / Proactive Q&A: ......

Most central: Papers 1, 3, 5, 6, 8 — they collectively explain why multi-turn formats are necessary, where they fail, how to evaluate them, and concrete mitigation strategies (including a domain example in finance). - Useful context: Papers 2, 4, 7, 9 — add signals ...... } ,
"step3_final_selection": { "final_selection": [ {"paper_index": 1,"rank": 1,"justification": "Frames clarifying questions as central to conversational and iterative Q&A, with evaluation evidence on simulators vs human reformulations; directly relevant to intent clarification across formats."}, {"paper_index": 3,"rank": 2,"justification": "Identifies multi-turn performance degradation and its causes, providing a theoretical and empirical foundation for why structured and iterative formats are needed to maintain prompt accuracy."}, ......] } }
\end{jsonbox}

\section{Prompt Templates}\label{sec:prompt}

In this appendix, we include the key prompt templates used in the SciRAG framework for different stages of the pipeline. These prompts are designed to ensure consistency, interpretability, and modularity across planning, retrieval, reasoning, and synthesis phases. 

\begin{tcolorbox}[colback=blue!5!white, colframe=blue!75!black, title=Outline Generation Prompt]
"Given a knowledge-intensive scientific question, please understand and analyze the question carefully. Consider key aspects like the intent behind the question, the motivations, potential flaws, and possible solutions. Based on this analysis, create a simple outline of the answer, specifying the parts and information that should be included in the response and the proportion each part should contribute. The total should add up to 100\%. The outline should guide what should be covered in the response, offering clarity on the scope and balance of each section."\\
"Your answer should be marked as [Response\_Start] Answer [Response\_End]."\\
"Here's an example outline:"\\
"Question: What strategies are used to improve robustness and safety of quadrotor UAVs in extreme weather conditions?"\\
"Answer: [Response\_Start]"\\
"1. (33\%) The answer should begin by explaining the importance of robustness and safety for quadrotor UAVs in extreme weather conditions."\\
"2. (33\%) The answer should discuss strategies and solutions to improve the robustness and safety of quadrotor UAVs in extreme weather conditions."\\
"3. (33\%) The answer should highlight the limitations or challenges associated with designing robust and safe solutions for quadrotor UAVs under extreme weather conditions."\\
"[Response\_End]"\\
"Now, please create an outline for this question: \{question\}"
\end{tcolorbox}

\begin{tcolorbox}[colback=blue!5!white, colframe=blue!75!black, title=Initial Answer Generation Prompt,fontupper=\small]
"Provide a detailed and informative answer to the following research-related question. Your response should offer a comprehensive overview and be clearly structured in multiple paragraphs."\\
"Organize your answer according to the key themes or sections identified in the outline below, and please note that the length of each part of the answer should be roughly the same as the percentage in the outline. Ensure each section is well-supported by multiple references, not just a single source."\\
"Focus on giving a comprehensive overview of the topic, rather than providing a short or surface-level response."\\
"Ensure the answer is well-structured, coherent and informative so that real-world scientists can gain a clear understanding of the subject. Rather than simply summarizing multiple papers one by one, try to organize your answers based on similarities and differences between papers."\\
"Make sure to add citations to all citation-worthy statements using the provided references (References). More specifically, add the citation number at the end of each relevant sentence e.g., 'This work shows the effectiveness of problem X [1].' when the passage [1] in References provides full support for the statement."\\
"Not all references may be relevant, so only cite those that directly support the statement."\\
"If multiple references support a statement, cite them together (e.g., [1][2]). Yet, for each citation-worthy statement, you only need to add at least one citation, so if multiple evidences support the statement, just add the most relevant citation to the sentence."\\
"Your answer should be accurate and rigorous, preferably with citations to support each sentence."\\
"References: \{context\}"\\
"Question: \{input\}"\\
"Outline: \{outline\}"\\
"Your answer should be marked as [Response\_Start] Answer [Response\_End]."
\end{tcolorbox}

\begin{tcolorbox}[colback=blue!5!white, colframe=blue!75!black, title=Gap Identification and Subquery Generation Prompts,fontupper=\small]
{\bf Prompt 1: Gap Identification}\\
"Please review the 'Current Answer' based on the 'Outline Guidance' and the 'Original Query'.\\
Your sole task is to accurately identify and describe what information, as required by the 'Outline Guidance', is missing from the 'Current Answer'. List these gaps clearly and specifically.\\
When reviewing, ignore any content related to future work, conclusions, or acknowledgments.\\
If the 'Current Answer' already fulfills all requirements in the 'Outline Guidance', state this explicitly. For example: 'The answer is complete and contains no information gaps.'\\
Outline Guidance: \{guidance\}\\
Current Answer: \{answer\}\\
Original Query: \{query\}"\\[1em]
{\bf Prompt 2: Decision and Subquery Generation}\\
"Please analyze the 'Identified Gaps' provided below.\\
If the 'Identified Gaps' list indicates that the answer is complete (e.g., the list is empty or explicitly states there are no gaps), you must return only '[end]terminate' in lowercase.\\
Otherwise, create one or more new search queries to gather the information needed to fill these gaps.\\
Follow these rules for creating queries:\\
- Merge Similar Queries: If multiple gaps can be addressed by similar queries, merge them.\\
- Minimize Query Count: Ensure each query explores a different sub-problem and provide as few queries as possible.\\
- Be Clear and Concise: Each query must be clear, concise, and contain necessary keywords to guide the retrieval process effectively.\\
- Do Not Reference the Answer: Do not mention or reference specific content from the 'Current Answer' in your new queries.\\
Please return your queries in the format below:\\
(1) Your query content.\\
(2) Additional query content if needed.\\
...\\
Identified Gaps: \{gap\_analysis\}\\
Original Query (for context): \{query\}"
\end{tcolorbox}

\begin{tcolorbox}[colback=blue!5!white, colframe=blue!75!black, title=Additional Answer Generation Prompt,fontupper=\small]
"You are in a retrieval chain that has been expanded to better answer the initial research-related core query."\\
"The retrieval path is: \{path\}."\\
"Currently, you are at the retrieval step for: \{query\}."\\
"Provide a detailed and informative answer only to the query at current step. Your response should offer a concrete answer."\\
"Make sure your answer includes summaries of relevant literature or texts or clear descriptions of their contribution to the query. When you make a claim, it is always best to have excerpts or citations to support them."\\
"Ensure your answer is well-supported by references. Focus on giving a concrete answer to the query, rather than providing a short or surface-level response."\\
"Ensure the answer is well-structured, coherent and informative so that real-world scientists can gain a clear understanding of the query. Rather than simply summarizing multiple papers one by one, try to organize your answers based on similarities and differences between papers."\\
"Make sure to add citations to all citation-worthy statements using the provided references (References). More specifically, add the citation number at the end of each relevant sentence e.g., 'This work shows the effectiveness of problem X [1].' when the passage [1] in References provides full support for the statement."\\
"Not all references may be relevant. You can read through the rationales and think on your own, and only cite those that directly support the statement."\\
"If multiple references support a statement, cite them together (e.g., [1][2]). Yet, for each citation-worthy statement, you only need to add at least one citation, so if multiple evidences support the statement, just add the most relevant citation to the sentence."\\
"Your answer should be accurate and rigorous, preferably with citations to support each sentence."\\
"References: \{context\}"\\
"Your answer should be marked as [Response\_Start] Answer [Response\_End]."
\end{tcolorbox}

\begin{tcolorbox}[colback=blue!5!white, colframe=blue!75!black, title=Reflective Refinement Prompts,fontupper=\small]
{\bf Prompt 1: Feedback Generation}\\
"Given an answer to a scientific question based on the most recent scientific literature, give me your feedback.\\
Ensure the answer is well-structured, coherent and informative so that real-world scientists can gain a clear understanding of the subject. Do not simply summarize multiple papers one by one, but you do should include proper summaries of papers, and try to organize your answers based on similarities and differences between papers.\\
Make sure your answer includes summaries of relevant literature or texts or clear descriptions of their contribution to the query. When you make a claim, it is always best to have excerpts and citations to support them.\\
Regarding the content improvements, it is often helpful to ask for more concrete results, applications, or methodologies to different tasks, elaborate on details of crucial methods, or suggest including explicit excerpts and citations as supports.\\
Stylistic improvements can include better organizations or writing enhancements.\\
Your answer should be marked as [Response\_Start] and [Response\_End].\\
If you think the current answer basically meets all the requirements and has no obvious room for improvement, and can be used as a candidate for a good answer, then return Feedback: [terminate] in lower case.\\
Else, prioritize the feedback by listing the most critical improvements first.\\
Each feedback should be preceded by 'Feedback: '.\\
The answer should be organized according to the outline below.\\
Question: \{question\}\\
Answer: \{answer\}\\
Outline: \{outline\}\\
\texttt{[Response\_Start]Feedback: [Response\_End]}\\
Now, please generate feedback for this question."\\[1em]

{\bf Prompt 2: Answer Refinement}\\
"You have been given a research-related question, an initial comprehensive answer, and some feedback pointing out possible improvements.\\
Now, please refine the answer according to the following guidelines:\\

1. \textbf{Focus and Organization}:\\
- Provide a thorough, multi-paragraph response, following the key themes or sections identified in the outline.\\
- Ensure that the approximate length and level of detail for each section is consistent with the proportions indicated in the outline, but don't include the percentage of the proportion in your answer.\\
- Rather than merely listing studies one by one, organize the discussion based on similarities or differences among the referenced works.\\

2. \textbf{References and Citations}:\\
- Use references from the 'References' section to support all citation-worthy statements, adding their citation number at the end of the sentence, e.g., '[1]'.\\
- If multiple references directly support the same statement, you may group them like '[1][2]'.\\
- Only cite references that truly support the claim, and ensure you re-index citations to match the final reference list if needed.\\
- Do not introduce references that are irrelevant to the statements being made.\\

3. \textbf{Clarity and Comprehensiveness}:\\
- Incorporate feedback to clarify or expand on crucial details, methods, or results.\\
- Strive for a more comprehensive overview of the topic rather than a surface-level summary.\\
- When making a claim or stating an important finding, it is best to briefly illustrate or quote relevant points from the supporting references.\\

4. \textbf{Feedback Integration}:\\
- Only modify parts of the original answer where the feedback indicates improvements are needed, keeping the other sentences unchanged.\\
- Do not omit any crucial information from the original answer unless the feedback explicitly states that certain sentences are incorrect or redundant and should be removed.\\
- If you add new paragraphs, ensure you are not duplicating content already present in the original response.\\

5. \textbf{Stylistic Consistency}:\\
- Keep the original paragraphs and new lines intact unless the feedback requires changes in structure.\\
- Maintain a coherent narrative flow, with smooth transitions between sections.\\
- Use clear, professional language that real-world scientists would find understandable and informative.\\

6. \textbf{Final Formatting}:\\
- Your refined answer must be enclosed between '[Response\_Start]' and '[Response\_End]'.\\
- Make sure the final version is well-structured, balanced according to the outline, and thoroughly addresses the question.\\

Below are the materials you have to work with:\\
- \textbf{Question}: \{question\}\\
- \textbf{Original Answer}: \{original\_answer\}\\
- \textbf{Feedback}: \{feedback\}\\
- \textbf{Outline}: \{outline\}\\
- \textbf{References}: \{references\}\\

Following these instructions, please refine the answer accordingly.\\
Your final answer should be marked between [Response\_Start] and [Response\_End]."
\end{tcolorbox}

\begin{tcolorbox}[colback=blue!5!white, colframe=blue!75!black, title=Branch-Based Synthesis Prompt,fontupper=\small]
"You are in a retrieval chain that has been expanded to better answer the initial research-related core query."\\
"The retrieval path is: \{path\}."\\
"Currently, you are at the retrieval step for query: \{query\}."\\
"Please review the current research-related query and its initial answer and read them carefully."\\
"The initial answer may have some shortcomings, so we performed additional searches and supplemented information. Now please combine the information from the supplemented query and answer to optimize the original answer, offering a comprehensive overview and clearly structured in multiple paragraphs."\\
"Also you should try to keep the original answer content's structure unchanged."\\
"Ensure the answer is well-structured, coherent and informative so that real-world scientists can gain a clear understanding of the subject, rather than providing a short or surface-level response."\\
"And re-cite the citations in the answer according to the latest reference list below."\\
"Make sure your answer includes summaries of relevant literature or texts or clear descriptions of their contribution to the query. When you make a claim, it is always best to have excerpts or citations to support them."\\
"Make sure to add citations to all citation-worthy statements using the provided references (References). More specifically, add the citation number at the end of each relevant sentence e.g., 'This work shows the effectiveness of problem X [1].' when the passage [1] in References provides full support for the statement."\\
"Not all references may be relevant. You can read through the rationales and think on your own, and only cite those that directly support the statement."\\
"If multiple references support a statement, cite them together (e.g., [1][2]). Yet, for each citation-worthy statement, you only need to add at least one citation, so if multiple evidences support the statement, just add the most relevant citation to the sentence."\\
"Your answer should be accurate and rigorous, preferably with citations to support each sentence."\\
"Here is the initial answer: \{answer\}"\\
"Here is the supplemented queries and answers: \{supplement\}"\\
"Here is the references: \{context\}"\\
"Your answer should be marked as [Response\_Start] Answer [Response\_End]."
\end{tcolorbox}

\begin{tcolorbox}[colback=blue!5!white, colframe=blue!75!black, title=System Instruction: Symbolic Reasoning Setup,fontupper=\small]
"You are in a retrieval chain that has been expanded to better answer the initial core query."\\
"The retrieval path is: \{path\}."\\
"Currently, you are at the retrieval step for: \{query\}."\\
"You have a set of partial paper texts (abstracts or snippets)."\\
"Your goal is to analyze each text's contribution and the relationship between them, build symbolic relationships,"\\
"and decide which texts are most relevant and contributing to the query and the overall chain."
\end{tcolorbox}

\begin{tcolorbox}[colback=blue!5!white, colframe=blue!75!black, title=Symbolic Reasoning — Step 1: Role Tagging and Relevance Assessment,fontupper=\small]
"We have the following candidate texts from different papers (abstracts or snippets): \{paper\_text\}"\\
"The query is: \{query\_text\}"\\

\textbf{Step 1 Task}:\\
1. For each paper, identify its key content segments and label them with:\\
\quad - T (theoretical part: theorem, definitions, main theoretical results),\\
\quad - E (experimental part: methodology, experiment details, results),\\
\quad - A (applications),\\
\quad - or other labels if needed (e.g., 'M' for methodology if it's not purely experimental).\\
2. For each segment, provide a brief summary (1–2 sentences)\\
\quad and assess its relevance to the query as High, Medium, or Low.\\
\textbf{Output format (example)}:\\
\{\\
\quad "papers": [\\
\quad\quad \{\\
\quad\quad\quad "paper\_index": 1,\\
\quad\quad\quad "segments": [\\
\quad\quad\quad\quad \{ "label": "T", "description": "...", "relevance": "High" \},\\
\quad\quad\quad\quad \{ "label": "E", "description": "...", "relevance": "Medium" \}\\
\quad\quad\quad ]\\
\quad\quad \},\\
\quad\quad ...\\
\quad ]\\
\}\\
Please keep the output structure strictly without additional comments.
\end{tcolorbox}

\begin{tcolorbox}[colback=blue!5!white, colframe=blue!75!black, title=Symbolic Reasoning — Step 2: Link Construction and Reasoning,fontupper=\normalsize]
"Below is the structured breakdown of each paper's segments from Step 1:"\\
\{step1\_result\_json\}\\
"Using that breakdown, please establish symbolic relationships among the papers and the query: \{query\}"\\
"   - T (theoretical part: theorem, definitions, main theoretical results),"\\
"   - E (experimental part: methodology, experiment details, results),"\\
"   - A (applications),"\\
"   - or other labels if needed (e.g., 'M' for methodology if it's not purely experimental)."\\
"For example:"\\
" - \"[1]T -> [2]T\" means paper1's theoretical part informs or extends paper2's theoretical part."\\
" - \"[1]E -> [Q]\" means paper1's experiment part contributes directly to answering the query."\\
" - \"[3]A -> [2]T\" means paper3's application part provides insights for paper2's theory."\\
"In each relationship, use the format: \"[paper\_index][label] -> [paper\_index or Q][label (if paper)]\"."\\
"If the second target is the query itself, just use [Q]."\\
\textbf{Output format (example)}:\\
\{\\
\quad "relationships": [\\
\quad\quad \{ "symbol": "[1]T -> [Q]", "rationale": "Paper1's theoretical result directly addresses the phenomenon in the query." \},\\
\quad\quad \{ "symbol": "[2]E -> [3]T", "rationale": "Paper2's experiment suggests data that confirms the theorem in Paper3." \},\\
\quad\quad ...\\
\quad ]\\
\}\\
Please keep the rationale concise, and keep the output structure strictly without additional comments.
\end{tcolorbox}

\begin{tcolorbox}[colback=blue!5!white, colframe=blue!75!black, title=Symbolic Reasoning — Step 3a: Coherence and Relevance Analysis,fontupper=\small]
"Given the symbolic relationships from Step 2 and the paper breakdowns from Step 1, your task is to perform a detailed analysis."\\
"Symbolic Relationships: \{step2\_relationships\_json\}"\\
"Paper Breakdowns: \{step1\_result\_json\}"\\
"Query: '\{query\}'"\\
"Retrieval Chain: '\{path\}'"\\
\textbf{Your Task:}\\
Analyze the coherence and relevance of the papers and their relationships in the context of the query. \textbf{Do NOT} decide which papers to keep or discard yet. Instead, provide step-by-step reasoning that addresses the following:\\
- \textbf{Identify Core Papers}: Which papers (and their segments like T, E, A) appear to be most central to answering the query? Explain why.\\
- \textbf{Identify Supporting Papers}: Which papers provide useful context or supplementary information but may not be essential?\\
- \textbf{Identify Contradictions or Weak Links}: Are there any relationships in the chain (e.g., T → T) that seem weak, irrelevant, or contradictory? For instance, does one paper's experiment invalidate another's theory?\\
- \textbf{Identify Irrelevant Papers}: Are there papers that seem entirely tangential or irrelevant to the specific query? Explain your reasoning.\\
Your output should be a clear, textual analysis that will be used in the next step to make final decisions.
\end{tcolorbox}

\begin{tcolorbox}[colback=blue!5!white, colframe=blue!75!black, title=Symbolic Reasoning — Step 3b: Final Decision and Ranking,fontupper=\small]
"We now have the symbolic relationships from Step 2:"\\
\{step2\_relationships\_json\}\\
Where we have the symbols:\\
\quad - T (theoretical part: theorem, definitions, main theoretical results),\\
\quad - E (experimental part: methodology, experiment details, results),\\
\quad - A (applications),\\
\quad - or other labels if needed (e.g., 'M' for methodology if it's not purely experimental).\\
And the breakdown of each paper from Step 1:\\
\{step1\_result\_json\}\\
For the query "\{query\}" within the context of the overall retrieval chain "\{path\}".\\
And based on the detailed 'Coherence and Relevance Analysis' provided below, your task is to make the final decisions on paper selection and ranking.\\
Coherence and Relevance Analysis: \{analysis\_from\_step3a\}\\
\textbf{Your Task:}\\
- \textbf{Finalize Selection}: Decide which papers to keep and which to discard based only on the provided analysis.\\
- \textbf{Rank Kept Papers}: Create a final ranked list (from most to least relevant) of the papers you decide to keep.\\
- \textbf{Justify Decisions}: For each kept paper, provide a concise justification for its rank. For each discarded paper, provide a brief reason for its exclusion. All justifications must be derived from the analysis. (You may consider the segments ([1]T, [1]E, etc.) internally, but your final output should only include paper indexes.)\\
\textbf{Output Format:}\\
You must return the final result in a valid JSON structure, exactly as shown in the example below, without any additional text or comments.\\
\{\\
\quad "final\_selection": [\\
\quad\quad \{ "paper\_index": 1, "rank": 1, "justification": "..." \},\\
\quad\quad \{ "paper\_index": 3, "rank": 2, "justification": "..." \}\\
\quad ],\\
\quad "discarded\_items": [\\
\quad\quad \{ "paper\_index": 2, "reason": "Not relevant to the query" \}\\
\quad ]\\
\}
\end{tcolorbox}

\section{Efficiency and Scaling Analysis}
We also report efficiency alongside scaling results on ScholarQA-CS here. 
Table~\ref{tab:efficiency} summarizes the average runtime and cost per query 
at different retrieval depths. As expected, deeper retrieval incurs higher 
time and monetary cost, but the growth remains manageable relative to the 
performance gains reported in the main text.

\begin{table}[h]
\centering
\small
\caption{Average runtime and cost per query at different retrieval depths (ScholarQA-CS).}
\label{tab:efficiency}
\begin{tabular}{@{}ccccc@{}}
\toprule
\textbf{Depth} & \textbf{Reasoning / Rerank Time} & \textbf{Other Components Time} & \textbf{Total Time} & \textbf{Total Cost (USD/query)} \\
\midrule
1 & 1m16s & 0m38s & 1m54s & 0.04 \\
2 & 1m52s & 0m47s & 2m39s & 0.09 \\
3 & 3m38s & 1m14s & 4m52s & 0.17 \\
4 & 5m13s & 1m59s & 7m12s & 0.29 \\
\bottomrule
\end{tabular}
\vskip -0.1in
\end{table}

\end{document}